\documentstyle[preprint,prb,aps]{revtex}
\begin{document}

%\twocolumn[\hsize\textwidth%
%\columnwidth\hsize\csname@twocolumnfalse\endcsname

\draft
\title{Island Nucleation in Silicon on Si(111) Growth 
under Chemical Vapor Deposition}
\author{K. E. Khor and S. Das Sarma} 
\address{Condensed Matter Theory Center, Department of Physics,
University of Maryland, College Park, Maryland 20742-4111}
\date{\today}
\maketitle

\begin{abstract}

Recent experiments show that the islanding behavior during chemical vapor
 deposition (CVD) of Si on Si(111) using disilane (Si${_2}$H${_6}$) is
quite different from that due to molecular beam epitaxy (MBE).
 While the latter can be understood using rate equation theories
(RET), the islanding exponent (connecting the power law growth of island density 
with growth rate)  obtained for the CVD growth is a puzzle, with the CVD exponent being almost 
twice the MBE exponent.
 We carry out (2+1) dimensional kinetic Monte Carlo(MC) simulations to study this CVD growth. 
 Hydrogen plays
a critical role during  growth. Disilane breaks up into hydrides 
on the Si surface. We use MC simulations to explore a number of
cases involving one or two migrating species and show that the large islanding exponent is probably
due to the presence of two hydrides, one of which has a much shorter lifetime than
the other. We modify RET taking this possibility into account in order to shed light on the
experimental observation.
We calculate the scaling properties of the island distributions using MC simulations and the modified RET,
and conclude that the large effective CVD exponents arise from the failure of the simple island 
number scaling scenario which no longer applies to the two-component situation prevailing under CVD
growth conditions.
\end{abstract}
%\vskip1pc]
\narrowtext
\section*{I. INTRODUCTION}
Molecular beam epitaxy (MBE) growth of semiconductors has been studied extensively
by means of rate equation theories; in particular, island nucleation and growth
of homoepitaxial silicon on silicon is known to be well described  by these 
theories \cite{and,ven,voi,mo}. On the other hand chemical vapor
deposition (CVD) growth has been much less well studied theoretically, even though it is widely used
in semiconductor technology. CVD homoepitaxy introduces atoms of one or more species
other than that of the substrate, which may alter substantially the physical and chemical 
properties of the surface, such as the surface free energy and the diffusion of atoms.
Clearly island nucleation behavior will also be affected under CVD conditions. 
In silicon homoepitaxy using disilane
Si$_2$H$_6$, the surface is hydrogenated as Si is deposited - the ratio (of Si to H) cannot be varied. 
It is known that the hydrogen atoms passivate the Si surface by saturating surface dangling bonds
In this study we look at the experimental work of Andersohn {\it et al}\cite{and}
on the CVD growth of Si on Si(111) using disilane gas. They obtained results for both MBE and CVD 
growth of Si/Si(111). From the perspective of kinetic Monte Carlo(MC) simulations, as we shall see later,
this CVD process is a straightforward extension of MBE; we can therefore, look for understanding 
of mechanisms which contribute to the differences between the two growth processes.
Andersohn {\it et al} obtained exponents for the island density $N$ versus growth rate $R'$
of 0.7 for MBE growth and a substantially larger value of  1.25 for CVD - in the latter
 case rate equation theories were 
applied naively. We study this system by means of MC simulations and  also generalize
and extend
standard rate equation theories to systems when more than one mobile species is involved
as is appropriate for CVD growth. Our goal is to understand
the substantial difference in growth rate exponents between MBE and CVD by carefully 
incorporating in the growth model the different adatom dynamics in the two cases.
The growth experiments of Andersohn {\it et al} were carried out in the temperature range
of 480$ ^{\circ}$C to 540$ ^{\circ}$C. Experimental work on disilane decomposition
on Si(111)7X7 surfaces suggests that incident disilane molecules interact with bare
(dangling bond) Si sites and deposit various hydrides such as SiH$_2$ and SiH
on the surface, with the latter being the only long-lived hydride\cite{kul}. The dihydride decomposes
to the monohydride and hydrogen; the hydrogen atoms may be left on the substrate to passivate  
dangling bonds. At the lower end of the above temperature growth range, the monohydride
breaks down very slowly through H evaporation or desorption. Once a site is passivated by hydrogen 
or a hydride, no further adsorption of disilane can happen at this site.    
STM images show that as a result of hydrogen passivation, the 7x7 surface 
reconstructions are replaced by the 1x1 H-stablised surface\cite{and,hoe}.
Kulkarni {\it et al} suggest that for temperatures above 490$ ^{\circ}$C, the
H desorption channel opens up rather substantially through removal by disilane molecules and also through the
recombination of hydrogen atoms from two nearest neighbor monohydrides, although
this latter process occurs at a much lower rate than the former. 
In the experiments of Andersohn {\it et al} Si is deposited on Si(111) at the rate of
1/10 to 10 bilayers(BL) per minute. The CVD experiments are
carried out in the submonolayer regime with Si coverages $\theta \leq 0.3$ monolayers(ML). 
   At 480$ ^{\circ}$C, the low end of the growth 
temperature range used, the hydrogen desorption rate is so small
that no significant amount of hydrogen has been removed during deposition. 
 This is important for CVD experiments, since the high growth exponents seem to apply to the
entire temperature range. This suggests that the large CVD growth exponents may be essentially independent
of hydrogen desorption.

 Andersohn {\it et al} plotted island density $N$ against growth rate $R$
for MBE growth for the two temperatures, 410$ ^{\circ}$C and 500$ ^{\circ}$C for a coverage
of $\theta \sim 0.3$ bilayers(BL); they fitted their results to the relationship given by
rate equation theories\cite{ven} (RET),
   
\begin{equation}
N=\left( \frac{R}{D_o} \right) ^\eta exp^{-\frac{E}{kT}},
\label{equ:e1}
\end{equation}

where $D_o$ is the diffusion preexponential and E an activation energy parameter, with the growth
or the islanding
exponent $\eta (= i^*/(i^*+2))$ for two dimensional (2D) islanding  for steady state nucleation
in the complete condensation regime\cite{ven}. In Eq.~\ref{equ:e1} $i^*$ is the critical 
cluster size for 2D islanding.
They obtained $\eta = 0.7$ for Si/Si MBE which corresponds to $i^* \sim 5-7 atoms$. 
For Si MBE where there is little or no evaporation, the deposition rate $R$ is equal to the
growth rate $R'$. Andersohn {\it et al} also applied Eq.~\ref{equ:e1} to their CVD experiments.
Mean growth rates $R'$ are calculated from the deposition times needed to attain a fixed coverage
of about a third of a bilayer
$\theta (\sim 0.3$ BL), which is determined from STM images so that material which has diffused
to step edges outside the imaged area is excluded from $\theta$. For CVD growth they obtained
exponents in the range $\eta \sim 1.1-1.25$, clearly substantially larger than the MBE
exponent $\eta \sim 0.7$. This is manifestly inconsistent with the above RET expression
 for $i^*$which asserts that $\eta$ must necessarily lie between 0 and 1.
The failure of the simple rate equation theory for CVD growth suggests that the presence of two
or more adatom species may alter nucleation behavior qualitatively, necessitating some new
insight into the RET formulation for CVD.  These species comprise
adatoms of Si and its various hydrides, SiH and SiH$_2$\cite{kul}. Of the hydrides, only SiH is
long-lived. The dihydride may decompose into a monohydride with the extra hydrogen
going to saturate a neighboring dangling bond on the Si(111) surface\cite{kul}.
 The monohydride should diffuse as a complex, since the Si-H bond is very strong
$\sim$ 3.1eV, compared to typical diffusion barriers of the order of 1eV or less\cite{kul,hoe}.
It is known that the mobility of the monohydride is much less than that of Si adatom
\cite{hoe,koh} in the experimental temperature range. It is reasonable to expect 
the mobility of the dihydride to be even less than that of SiH. We should also be able 
to assume that SiH$_2$ moves as a complex, again based on the Si-H bonding energies
being much larger than surface diffusion activation energies.  Kandel\cite{kan} has 
suggested a different mechanism involving island edge barriers, which we shall discuss later. 

  Andersohn {\it et al} discuss the possible causes for the anomalously high CVD exponents. They
consider various physical possibilities which may affect CVD islanding such as
  a strong growth rate dependent diffusion parameter $D_o$, the
transformation of the surface reconstruction from (7x7) to (1x1), possible release of
additional Si atoms from the substrate to form extra islands, and hydrogen etching of the surface,
introducing the possibility of incomplete condensation to the nucleation process.
In the current work we study the CVD problem by using kinetic Monte Carlo  simulations
and also by rate equation analysis. The MC simulation will be done in (2+1) dimensions
on a square lattice (the particular lattice should not matter since our goal is to understand
qualitatively the origin of the large exponents and not to obtain quantitative agreement 
with experiment). Activation barriers for diffusion in our simulation are chosen using
previous MC simulations for Si MBE growth as a guide\cite{kho} and also so as to obtain a maximum
attainable (in terms of reasonable simulation times and problem sizes) stable critical  
island size of $i^* =3$ (for the cubic lattice) for no-evaporation MBE growth. 
Then larger exponents, if they do arise, say, in the CVD case cannot be attributed to regular causes.
Standard rate equation theories assume a single migrating species, usually lone adatoms. In our 
simulations for CVD we will consider one or more diffusing species composed of Si adatoms
and the silicon hydride complexes. We will modify rate equation theories to handle the situation
with two species.

 We carry out MC simulations for the following cases:\\
\indent(1)Si MBE without evaporation,\\
and several distinct situations for CVD growth as given below:\\
\indent(2)migrating species of SiH, with H allowed to evaporate and the resulting
lone Si atom allowed to migrate rapidly to a step edge;\\
\indent(3)two migrating species of Si and SiH, with H evaporation allowed;\\
\indent(4)deposition of SiH at a site is accompanied with an extra H saturating a neighboring 
dangling bond; this represents the decomposition of SiH$_2$ discussed above,\\
and finally;\\
\indent(5)two migrating species, comprising SiH$_2$ and SiH, with the former being much 
slower and short-lived.\\ 
The details for the MC simulation model for these cases are presented in section II. In  
section III we present and discuss the simulation results and provide a critical comparison
between the MBE and the CVD case. We conclude in section IV summarising our results and discussing
the extended RET model applied to this problem. In appendix A  we give the details for
the generalized rate equation theory appropriate to the two species system. 
Appendix B gives the standard RET in terms of the
actual growth rate rate $R'$ (rather than the deposition rate R ).

\section*{II. KINETIC MONTE CARLO SIMULATION}
MC simulations are carried out in (2+1) dimensions on a square lattice. An adatom or adatom complex
moves under the solid-on-solid restriction (overhangs and/or bulk vacancies are not allowed). 
These simplifications 
are not expected to be important for the study of growth exponents.
An adatom moves to a neighboring site by randomly hopping at a rate that depends on its 
bonding configuration; this rate is given by 

\begin{equation}
R_n=R_o exp^{-\frac{E}{kT}},
\label{equ:e2}
\end{equation}

where $R_o=d'kt/h$ is a characteristic vibrational frequency and d'(=2, in our case) is 
the substrate dimension. $E=E_o+nE_b$ is the activation energy, where n is the number of
nearest neighbors and $E_b$ is the energy (actually the activation barrier) associated with
 the bond between nearest neighbors. Since we are
dealing with up to two different species comprising adatom complexes from Si, SiH and SiH$_2$,
the parameter $E_o$ will be different for different species. Although this will mean that
each species will diffuse at different rates, we will assume that when two adatom complexes
are next to one another, the nearest neighbor bonding remains the same irrespective of their
species. This seems to be a reasonable assumption since the bonding is between the Si atoms
of the hydrides, and the hydrogen atom is not expected to affect this much. Even if there is some
species dependence of this nearest neighbor bonding or activation energy, this
should not affect our qualitative conclusion.  We will also
assume that the hydride SiH (and SiH$_2$) will hop as a complex since the bond between Si and
H atoms is extremely strong $\sim 3.1$eV\cite{kul,hoe} compared to the  hopping activation barriers
$\sim 1$eV. These assumptions are supported by the experiments\cite{kul} of Kulkarni {\it et al}  on the 
decomposition of disilane on Si(111). In our simulations we allow for the evaporation of hydrogen
through a single evaporation rate for the sake of simplicity. Kulkarni {\it et al}\cite{kul}
  describe two possible
mechanisms for H removal, but only one is significant in our context. 
We do not expect the simplification of 
 using just one H-evaporation rate to affect our conclusions. In accordance with experimental 
observations\cite{and,kul} once a site is passivated by a silicon hydride complex, no
further deposition on that site is possible; correspondingly an adatom complex can only
jump to a H free site.  

Our simulation parameters, together with the deposition and evaporation rates, have been chosen so that
a critical island size of $i^*=3$ is obtained for MBE growth in the temperature ranges 
of interest. Results from standard nucleation theory\cite{lew} suggest that, for a square
lattice as the stable critical island size increases from $i^*=1$ to 3 (and then to 10), 
the deposition rate
decreases by two orders of magnitude for each step in the increase of $i^*$ (i.e. R has
to decrease by a factor of $10^4$ if $i^*$ increases from 1 to 10). We use deposition
rates in the  range $0.3 \leq R \leq 32$ monolayers(ML)/s. For MBE we obtain straight line plots of
ln$N$ versus ln$R$, indicating a power law island density growth as a function of
the deposition rate, in agreement with Eq.~\ref{equ:e1}. At the lower limit of $R \sim 0.3$, simulation
times become so large that it is unlikely that calculations can be extended  much below this limit. 
Exponents larger than $\eta =0.6$, (i.e. $i^*=3$)  for MBE are therefore beyond reasonable 
simulation times,which is the usual finite time problem in kinetic simulation of very slow
dynamics. Note that $\eta <1$, satisfying the theoretical RET bound of $0<\eta<1$ corresponding
to $0<i^*<\infty$ for the island size.  Larger exponents (in the CVD simulations) 
have to be attributed to mechanisms not operational in MBE growth.

We explore a variety of growth models. Simple MBE of Si on Si(111) provides a starting
reference point. Next, since H plays a critical role in CVD growth and the monohydride is
the only long-lived species on the substrate, we study a model where SiH is deposited but
that the hydrogen is allowed to evaporate at certain rates; if this results in a lone 
Si adatom, it moves instantaneously to the step edge and is lost to the island nucleation
process, as seen in the experiments of Andersohn {\it et al}. Si adatoms are much more
mobile than the hydrides\cite{and,hoe}. Kulkarni {\it et al} discuss a deposition process
where a dihydride breaks up into a monohydride and a hydrogen which passivates a neighboring
dangling bond. We explore the effect of this process on nucleation to see
if it would lead to an effective deposition rate dependent diffusion of SiH. Finally, we
look at the situation with two active migrating species. We assume that of the hydrides the dihydride
moves much more slowly than the monohydride.    
Another example of growth under two migrating species is to consider the combination of
 Si and SiH complexes, with the former being much faster than the latter. However, we do 
not find exponents markedly different from the MBE case until almost all the H has evaporated from
the hydride, which is experimentally not the case.  For this reason we will not discuss this
case further. In the next section we present and discuss our numerical results based on the 
kinetic MC simulations of Si/Si MBE and CVD process.

\section*{III. RESULTS}
To model MBE of Si on Si(111) we choose the set of parameters, $E_o=1.0eV$ and $E_b=0.3eV$.
Simulations are carried out on system sizes of 320x320; results remain essentially the same
when the size is increased to 480x480.
In Fig. \ref{fig1} we plot island densities $N$ versus deposition rates $R$ for the
growth temperature T=780K. The solid curve shows results for MBE growth. We count
islands comprising two or more atoms as stable; this lower limit of 2 does not seem to affect
the numerical value of the growth exponent we get in our simulations. Using 3-5 atoms
 as the lower limit on the stable island size gives essentially the same results. We
obtain an MBE growth exponent $\eta=0.6$, corresponding to a critical island nucleus of $i^*=3$. Over
nearly two orders of magnitude in the deposition rate $R$, the ln(N)-ln(R) plot in
 Fig. \ref{fig1} is essentially a straight line,
as discussed above. We also look at the distribution $N_s(\theta)$ of island size s.
In Fig. \ref{fig2} we display scaled island size distributions for two deposition rates 
$R=1$ and 16 and coverage $\theta \sim 0.095$. The lines through points (squares, diamonds
and circles) are due to a scaling ansatz proposed by Amar and Family\cite{ama},       

\begin{equation}
\frac{N_s(\theta )S^2}{\theta}=f_i (\frac{s}{S}),
\label{equ:e3}
\end{equation}

where $S(\theta )=\Sigma_s sN_s(\theta )/\Sigma N_s(\theta )$ is the average island size
and $f_i(u)=C_iu^iexp(-a_iu^{1/a_i})$ is  a scaling function, with 
$a_i$ satisfying the relation $\Gamma [(i+2)a_i]/(\Gamma [(i+1)a_i]=(ia^i)^{a_i},
 C_i=(ia_i)^{(i+1)a_i}/(a_i\Gamma [(i+1)a_i]$
and $i$ is the critical island size. We see that $i \leq i^*(=3)$ and is not constant over the
range of $\theta $ investigated, which suggests that critical sizes obtained from fitting island 
distributions with the scaling relation defined by Eq.~\ref{equ:e3} is less than those given
 by rate equations.
In our simulations, in contrast to the work of Amar and Family\cite{ama}, we do not impose the unnecessary
constraint that atoms with $i$ or more neighbors are not allowed to detach from an existing
island.

In the second set of simulations, SiH is deposited on the substrate at various rates;
the hydride is allowed to move as a complex with a hopping rate defined 
by the activation energy $E_o=1.0eV$.
The hydrogen evaporates at certain rates, leaving bare Si atoms. Lone Si atoms diffuse 
fast enough so that they are immediately lost to a step edge ( i.e. the atom
is effectively removed). This is consistent with the manner in which the growth rate is estimated
in the experiments of ref. 1. In Fig. \ref{fig1} we display results of 
simulations for two evaporation rates $R_e$=1200 and 2000. Both curves approach the MBE
results at high deposition rates $R$ and show higher slopes at lower $R$ indicating
higher rates of CVD exponents(at least) for lower deposition rates. These higher
exponents are consistent with rate equation theory results for incomplete condensation\cite{ven}.
However, the exponents of Andersohn {\it et al} are obtained from plots of island density $N$
versus the effective growth rate $R'$, not the deposition rate $R$. The growth rate 
$R'$ is estimated from the accumulating  material that stays
within the area imaged by the STM used in data analysis. When our results are
 replotted against $R'$, the two curves
rescale to the MBE results for all growth rates, and the effective CVD exponent for $N(R')$
becomes the same as the MBE exponent for $N(R)$. This effect can also be seen in rate
equation theories which we discuss in Appendix B. 

Andersohn {\it et al}\cite{and} discuss the possibility of a strong growth rate dependence of 
 the diffusion parameter $D_o$, due to the presence of hydrogen. Kulkarni {\it et al}\cite{kul} show
that among the various surface chemical reactions that can happen, there is one where a silicon dihydride
complex breaks up into a monohydride complex and a hydrogen which saturates a neighboring 
dangling bond; at low coverages, this dangling bond is likely to be associated with
a substrate Si atom with four Si nearest neighbors, which would make {\it this} particular SiH complex
essentially immobile. It is possible that this H atom, produced by the local chemical reaction
SiH$_2 \rightarrow$ SiH+H,  will reduce the effective mobility 
of a nearby mobile SiH complex. We test this idea in our simulations by depositing SiH$_2$
which breaks up immediately into a monohydride and a hydrogen atom.  This atom passivates a
neighboring dangling bond site chosen at random within 2 lattice spacings from the
monohydride; all lone monohydrides are allowed to move with hopping rates determined by 
the parameters above, no hydrogen evaporation is considered. We show these modified simulation
 results also in
Fig. \ref{fig1}. The curves are parallel to the line for MBE; they are shifted towards
higher densities, giving the same value( $\sim$ 0.6) for the exponent $\eta $. It is as if the H
has increased the effective deposition rate by a fixed percentage for all growth rates without
in any way affecting the underlying island scaling behavior.

Andersohn {\it et al}\cite{and} has noted that the presence of H changes the 7x7 reconstruction
of the surface to a 1x1 H-terminated surface which may result in up to 20\% of the 
material in island nuclei coming from the substrate itself. As noted above the result
of increasing the effective deposition rate by a fixed percentage simply raises the
curve by a constant amount without changing the growth exponent. Thus, this mechanism should not affect
the exponent in any significant manner.        

Finally we simulate growth due to the explicit presence of two species on the substrate.
Kulkarni {\it et al}\cite{kul} show that above 250$ ^{\circ}$C, disilane decomposes to leave
SiH$_2$ on the surface. This dihydride in turn breaks down to leave the monohydride plus
hydrogen which may or may not desorb. Since this hydrogen does not affect the growth exponent, as
we discuss above, we will neglect it in our simulation and simply allow it to desorb.
  Consider the two adatom species: SiH$_2$
which is first deposited has a very short lifetime and so we take SiH to be infinitely long-lived.
We use the same bonding strength $E_n=0.3eV$ for both and $E_o=1.0eV$ for the diffusion of
SiH; since SiH$_2$ is very much less mobile than SiH, we take $E_o$ for SiH$_2$ to be 2eV. The 
breakup rate of SiH$_2$ is given by setting $R_e$=730, chosen to optimize the exponent.
Clearly, this parameter $R_e$ sets the proportion of SiH to SiH$_2$ on the substrate in the steady state.
 The results are not sensitive to this particular value of $R_e$. The experimental evidence
is overwhelming that hydrogen is essential in maintaining growth with the high
 exponent value \cite{and,hoe}. If
the flux of disilane is low or growth occurs at high temperatures so that hydrogen desorption
is fast enough to keep the surface essentially H free, then the CVD growth
 becomes essentially the same as the 
MBE growth in terms of the value of the islanding exponent. For this reason and due to the fact
 that large CVD exponents are
seen in growth at 480$ ^{\circ}$C and 500$ ^{\circ}$C when H desorption is 
insignificant\cite{and}, we ignore the loss of H by SiH itself in our simulations.   
In Fig. \ref{fig1}, we plot the results of $N_s(\theta )$ versus deposition rate for this case
of 2-species growth. The exponent
$\eta=0.9$ is higher than that for MBE of 0.6 (which corresponds to $i^*=3$). On the basis
of the old rate equation theories an exponent of 0.9 corresponds to a critical nucleus
of $i^*=20$, which is greater than the size of the majority of the islands for the high
deposition rates, $R \geq 10$. In Fig. \ref{fig3}a, the island distribution results averaged
over 40-50 samples, for $R=1$ are displayed, together with the scaling relations 
for $i^*=1,2$ and 3. Clearly the results do not really fit any one relation but are closest
to the $i^*=2$ curve for $s/S \geq 1.5$, while they are closest to the $i^*=1$ curve at
$s/S \sim 1$; here the implication is that $i \sim 1$ does not agree with $i^* \sim 20$
as suggested by rate equation theories. In Fig. \ref{fig3}b, the corresponding curve for $R=16$
shows no agreement with any of the scaling relationsips at all. Island density declines
monotonically to a tail longer than those due to scaling. Thus, in the presence of 2-species
migration-induced islanding and growth, simple scaling relations break down and the effective
scaling exponent $\eta$ becomes very large indicating a failure of the simple scaling relation.

\section*{IV. DISCUSSION AND CONCLUSION}
Rate equation theories show that for 2 and 3D islanding, growth exponents of $
\eta \geq 1$ occur in the steady state case, only under regimes of extreme incomplete 
and initially incomplete condensation\cite{ven}. Our kinetic MC simulations, including those
not covered by standard rate equation theories, indicate that it is not simple to obtain 
these high growth exponents from naive CVD simulations. Even in the case which 
involves evaporation of H, the high
exponent at low effective growth rate reduces to the low exponent case when island density is plotted
against growth rate $R'$ instead of deposition rate $R$. As we will see in  Appendix B
the case of initially incomplete condensation collapses into that of complete 
condensation when the island density is plotted against growth rate $R'$; the exponent
 $\eta = i/(i+2.5), i/(i+2)$ for 3D and 2D islanding respectively\cite{ven}. In our
simulations, we explored several physical processes and mechanisms suggested by the CVD experiments.
 We find that the
mechanism of the change of surface reconstruction from 7x7 to 1x1, which can increase
the effective deposition rate by a given percentage, cannot increase the exponent.    
Andersohn {\it et al}\cite{and} suggest that a monomer diffusion constant that is strongly $R$
dependent could account for the high exponent; we explored a possible way - the 
codeposition of H in the neighborhood of a deposited monohydride - and found that this
merely raised island densities without increasing $\eta $. It is interesting that this
has the same effect as increasing the deposition rate by a fixed percentage. 
%has the same effect of increasing the deposition rate by a fixed percentage. ------
Based on the investigations\cite{kul} of Kulkarni {\it et al}, who showed that at the growth temperatures
of ref. 1 disilane deposits a dihydride on the substrate which subsequently  
breaks up into H and the monohydride, we investigate a growth model involving these two
hydrides ("the 2-species migration model" for CVD). The important 
characteristics of this model are that each hydride is assumed to
 move as a complex, since the SiH bond is known to be very
strong and that the dihydride, which moves much more slowly than the monohydride, has a
much shorter lifetime. In our simulations of this two component system an effective growth exponent
$\eta$ 50\% higher than the corresponding MBE case ($\eta \approx 0.6$) is obtained.
 We find $\eta =0.9$ for the square
lattice 2-component CVD growth and expect to see still higher exponents for
the realistic diamond lattice where the smallest stable
clusters of size greater than three would be accessible to simulations within reasonable
times. Our conclusion is, therefore, that this 2-component model of 2-species migration
is operational in the CVD growth of ref.1, leading to very large CVD growth exponents.

In Appendix A we discuss the extension of Walton's expression for the equilibrium density
of critical clusters\cite{wal} to the two component system that we study here. We then
extend standard rate equation theories\cite{ven} to these systems
and show that the exponent goes as $i/2$ instead of $i/(i+2)$ for 2D islanding and
$2i/5$ instead of $i/(i+2.5)$ for 3D islanding.
Large growth exponents have also been shown to arise when island edge barriers $E_B $ are
present\cite{kan}. Kandel suggests\cite{kan} that these edge barriers may occur in surfactant
mediated growth; an adatom can only be attached to the edge of an island after surfactant
atoms have been removed. The energy barrier required for such an interchange may be very large.
Kandel argues that the H in disilane CVD on Si plays the role of surfactant atoms. Kandel's
work has been discussed  and criticised in some detail by Venables and Brune\cite{bru}; they showed that
this island edge barrier has significant consequences: Even a modest edge barrier would cause
the steady state nucleation stage when the population of monomers $n_1$
has become constant, to be greatly delayed, that is, the transient stage ($n_1=R(1-Z)t= \theta (1-Z)
$) will become substantially extended.  Here Z is the coverage due to clusters excluding monomers.
This is because the capture numbers $\sigma_1, \sigma_x $, which determine the rate of removal
of monomers by clusters of size 1 and $x$ respectively, are reduced by 
the factor $exp-\beta E_B$ and, therefore,
 nucleation times are enormously increased by
the inverse of this factor. Our estimates are that, with the parameters we use in our 
simulations, a barrier of $E_B \sim $ 1eV  is enough to move the transient regime up into
the range of 
coverages of the order of 0.1ML. Venables and Brune conclude that the exponent of 
$\eta =2i/(i+3) $ obtained by Kandel, based on steady-state results may not be valid for 
his comparisons with experiments\cite{kan}. Furthermore, we examine the assumption that H
behaves as a surfactant. Kulkarni {\it et al} have shown\cite{kul} that Si-H moves as a complex; as
we have noted before, the Si-H bond is very strong\cite{hoe}, so that the H atom does not
behave like a surfactant. There is no exchange of Si with hydrogen for the silicon atom to
join an island at the edge of the island. It is known that once hydrogen passivates a 
dangling bond on the Si surface, there can be no further growth on top\cite{and}.
Our conclusion based on our kinetic MC simulations is that the large CVD exponents reported
in ref.1 arise from a failure of the simple island number scaling scenario in the 2-component
 migration condition prevailing under CVD growth. More experiments and simulations
are obviously needed to settle the issue definitively.

This work is supported by the US-ONR and NSF-MRSEC.

\appendix
\section{ TWO COMPONENT RATE EQUATIONS}
We consider a system where only monomers are mobile. We have 
a two component system with monomer densities $n_{1H}$ and $n_{2H}$ and total
density of monomers $n_1=n_{1H}+n_{2H}$. These two types of monomers move with diffusion
constants $D_{1H}$ and $D_{2H}$, but in forming islands the monomers are linked to each other by bonds
of the same strength $E_b$, so that clusters of size i have energies $E_i$, which are
independent of the detailed composition of adatom complexes of type 1H and 2H. Under these
conditions Walton's equation\cite{wal} for the equilibrium density , $n_i$, of clusters of size i,
given by 
 
\begin{equation}
\frac{n_i }{N_o}=C_i (\frac{n_1}{N_o})^i exp(\frac{E_i}{kT}),
\label{equ:A1}
\end{equation}

should apply. Here $E_i$ is the energy of the cluster relative to the monomer state, $N_o$
is the site density, and $C_i$ is a weighting factor dependent on the lattice.
Only monomers of type 2H are 
deposited; they have a lifetime of $\tau_2 $ and decay to type 1H monomers. Following
Venables\cite{ven}, we write down the rate equations for this two component system,

\begin{equation}
\frac{dn_{2H}}{dt}=R(1-Z)-\frac{n_{2H}}{\tau_2}-\sigma_{2x}D_{2H}n_xn_{2H},
\label{equ:A2}
\end{equation}

where $R$ is the deposition rate, $Z=a_xn_x$ is the coverage due to clusters (whose 
average area is $a_x$), $\sigma_{2x}$
is the average capture number by clusters of the 2H monomer and $n_x$ is the density of 
stable clusters. 

\begin{equation}
\frac{dn_{1H}}{dt}=\frac{n_{2H}}{\tau_2}-\sigma_{1x}D_{1H}n_xn_{1H},
\label{equ:A3}
\end{equation}

where $\sigma_{1x}$ is the average capture number by clusters of the 1H monomer. From Eqs.~\ref{equ:A2}
and ~\ref{equ:A3}, we obtain 

\begin{eqnarray}
\nonumber \frac{dn_{1}}{dt}&=&R(1-Z)-\sigma_{1x}D_{1H}n_xn_{1H}-\sigma_{2x}D_{2H}n_xn_{2H}\\
                 &=&R(1-Z)-\sigma_{x} n_{1}\left\{\frac{D_{1H}n_{1H}}{n_{1}}+
                \frac{\sigma_{2x}}{\sigma_{x}}\frac{D_{2H}n_{2H}}{n_{1}} \right\},
\label{equ:A4}
\end{eqnarray}

where we have written $\sigma_{x}$ in place of $\sigma_{1x}$ for notational simplicity.
 We can write the second term of
Eq.~\ref{equ:A4} as $\sigma_{x} n_{1}D$, where D is an effective diffusion constant. The equation for
stable clusters is given by

\begin{equation}
\frac{dn_{x}}{dt}=U_{i}-U_{c},
\label{equ:A5}
\end{equation}

where $U_{i}$ is the nucleation rate of stable clusters (i.e. clusters with size $>i$) 
and $U_{c}$ is the rate at which stable
clusters coalesce. $U_{i}$ is given by

\begin{eqnarray}
\nonumber U_{i}&=&\sigma_{1i}D_{1H}n_in_{1H}+\sigma_{2i}D_{2H}n_in_{2H}\\
\nonumber                 &=&\sigma_{i} n_{1}n_{i}\left\{\frac{D_{1H}n_{1H}}{n_{1}}+
                \frac{\sigma_{2i}}{\sigma_{i}}\frac{D_{2H}n_{2H}}{n_{1}} \right\}\\
                 &=&\sigma_{i} n_{1}n_{i}D,
\label{equ:A6}
\end{eqnarray}
 
where we have written $\sigma_{1i}$ as $\sigma_{i}$ and the effective diffusion constant as the
same D as in Eq.~\ref{equ:A4}. This approximation is consistent with the one we will make below.  We use
an expression for $U_{c}$ given by $U_{c}=2n_{x}dZ/dt$ \cite{ven}. Eqs.~\ref{equ:A4}, 
~\ref{equ:A5} and ~\ref{equ:A6} are now 
identical to those in Venables \cite{ven}. Together with the 2D islanding equation for $Z$,
$dZ/dt=\Omega^{2/3}(\sigma_{x}Dn_{1}n_{x}+RZ)$,and Walton's equation (Eq.~\ref{equ:A1}) they lead to the
 following equations\cite{ven},

\begin{equation}
\frac{dn_{1}}{dt}=R(1-Z)-\frac{n_{1}}{\tau},
\label{equ:A7a}
\end{equation}
 
where $1/\tau=\sigma_{x}Dn_{x}$ and

\begin{equation}
\frac{dn_{x}}{dt}=\frac{\gamma_{i} N_{o}^{1-i}Dn_{1}^{i+1}}{(1-Z)^{i}}-2n_{x}\frac{dZ}{dt},
\label{equ:A7b}
\end{equation}

where $\gamma_{i}=\sigma_{i} C_{i} exp(\beta E_{i})$ and $N_{o}$ is the number\cite{ven} of sites/$cm^{2}$.
Under steady state conditions Eq.~\ref{equ:A7a} gives $n_{1}=R \tau (1-Z)$. Using this result and the variables 
$\mathcal{N}$ = $n_{x}/N_{o}$ and $\mathcal{T}$ = $Rt/N_{o}$, Eq.~\ref{equ:A7b}  becomes

\begin{equation}
\frac{d \mathcal{N}}{dZ}=  (1-Z) \gamma_{i} (\frac{R}{DN_o^{2}})^{i}
 \frac{1}{ N_{o} \Omega^{ \frac{2}{3}}}(\frac{1}{\sigma_{x} \mathcal{N}})^{i+1} -2\mathcal{N},
%( \sigma_{x} N )^{-i-1} \frac{1}{ N_{o} \Omega^{ \frac{2}{3}}}-2Z,
\label{equ:A8}
\end{equation}

where we have used the result

\begin{eqnarray}
\nonumber  \frac{d \mathcal{T}}{dt}&=& \frac{R}{N_{o}} \frac{1}{dZ/dt} \\
                        &=& \frac{1}{N_{o} \Omega^{\frac{2}{3}}}.
\label{equ:A9}
\end{eqnarray}

Under steady state conditions, we obtain from Eq.~\ref{equ:A3} the result 
$n_{1H}/n_{2H}=1/\alpha$, where $\alpha = \tau_{2} \sigma_{x}D_{1H}n_{x} $.
This gives $n_{1H}/n_{1}=1/(1+\alpha)$ and $n_{2H}/n_{1} = \alpha/(1+\alpha)$
and if we assume  that $\sigma_{2x} \sim \sigma_{x}$, then the 
effective diffusion constant in Eq.~\ref{equ:A4} is given by 
 $D=(D_{1H}+ \alpha D_{2H})/(1+ \alpha)$. We now 
consider two limits; first,(a) when $D_{1H} \ll \alpha D_{2H}$,  $D=D_{2H} \alpha/(1+ \alpha)$.
Here a growth exponent other than the standard one results when $\alpha < 1$. This leads to
$\eta = i/(2i+3)$. However the diffusion
of the dihydride is much slower than that of the monohydride, so that the other limit (b)
$D_{1H} \gg \alpha D_{2H}$ is applicable; then $D$ is well approximated by $D=D_{1H}/(1+ \alpha)$.
Here there are two limiting cases that we can look at; the first is  
where the second term in the denominator is less than 1 and the second when it is greater than 1.
The first case simply leads to the standard growth exponent for 2D islands for the complete
condensation regime\cite{ven1}, $\eta=i/(i+2)$. The second case may apply to the experimental
 results of Andersohn {\it et al}. Using the parameters\cite{and,ven}, $n_{x} \sim 10^{12}/cm^{2}$, 
$1<\sigma_x <100$ and estimating from the results of Mo {\it et al}\cite{mo} that
$10^{-10}<D_{1H}<10^{-8}(cm^{2}/sec)$ at 500$ ^{\circ}$C and assuming that $\tau_{2}$  is at 
least of the order of 1 second, we can see that the second term can be dominant. In this
case Eq.~\ref{equ:A8} becomes      

\begin{equation}
\frac{d \mathcal{N}}{dZ}=  (1-Z) \gamma_{i} ( \frac{R \tau_{2} }{N_{o}})^{i}
\frac{1} { \sigma_{x} \mathcal{N}}  \frac{1}{N_{o} \Omega^{ \frac{2}{3}}}-2\mathcal{N},
\label{equ:A10}
\end{equation}
 
Setting $ \mathcal{N}$ = $\zeta (Z) \left( \gamma_{i} \frac{( R \tau_{2})^{i}}{N_o^{i+1}
 \Omega^{ \frac{2}{3}}} \right) ^{ \frac{1}{2}}$,
where $ \zeta(Z)$ is a pure number satisfying a differential equation\cite{ven}. This 
leads to the growth exponent $\eta = i/2$ for 2D islanding. A similar analysis for 3D islanding
leads to $\eta = 2i/5$. In both cases the exponent could be larger than unity.

\section{ RATE EQUATIONS IN TERMS OF GROWTH RATE $R'$}
We look at the standard rate equations in terms of the growth rate $R'$ rather than the
deposition rate $R$. Eqs.~\ref{equ:A7a},~\ref{equ:A7b} remain the same but $\tau$ is now given by   
$1/\tau=1/\tau_{a}+\sigma_x Dn_x$ , where evaporation is now considered and
$\tau_a$ is the evaporation lifetime of the monomer. Eq.~\ref{equ:A9} becomes

\begin{equation}
\frac{d\mathcal{T} }{dt}=\frac{1}{N_o \Omega^{\frac{2}{3}}} \delta,
\label{equ:B1}
\end{equation}

where $A=D \tau_a N_o$ and $\delta=( 1+\sigma_{x} A \mathcal{N})/ 
(\mathit{Z+\sigma_{x} A \mathcal{N}})$. Eq.~\ref{equ:A8}  becomes

\begin{equation}
\frac{d \mathcal{N} }{dZ}= (1-Z) \sigma_{i} B_i(\frac{A}
{1+\sigma_{x} A \mathcal{N} })^{i+1}   \frac{d \mathcal{T} }{dZ}-2\mathcal{N} ,
\label{equ:B2}
\end{equation}

where $B_{i}=C_{i}exp(\beta E_{i}) (\frac{R}{N_{o}^{2}D})^{i}$.
 From $R'=R-n_1/\tau_a $ and the steady
state relation $n_1=R \tau (1-Z)$, we get $R=R' \delta$.  In terms of $R'$, 
$B_i=B_i' \delta^i$, where $B_i'=C_iexp(\beta E_i) (\frac{R'}{N_o^2D})^{i}$.
We obtain growth exponents for the following condensation regimes:\\
(a)for extreme incomplete condensation $\sigma_x A \mathcal{N} \ll \mathit{Z}$. 
This leads to $R=R'/Z$
and $B=B'/Z$. The growth exponent plotted against $R'$ remains the same as that plotted
against $R$, with $\eta =i$.\\
(b)for complete condensation $\sigma_x A \mathcal{N} \gg \mathit{Z}$. This gives $R=R'$ and $B=B'$ and
so the exponent remains unchanged at $\eta =i/(i+2)$.\\
(c)for initially incomplete condensation $1 > \sigma_x A \mathcal{N} > \mathit{Z}$. We obtain

\begin{equation}
\frac{d\mathcal{T}}{dt}=\frac{1}{N_o \Omega^{\frac{2}{3}}}  \frac{1}{\sigma_x A \mathcal{N}},
\label{equ:B3}
\end{equation}

\begin{equation}
B_i=B_i' (\frac{1}{\sigma_x A \mathcal{N}})^i,
\label{equ:B4}
\end{equation}

\begin{equation}
\frac{d \mathcal{N}}{dZ}=  (1-Z) \sigma_i B_i' ( \frac{A}
{ \sigma_x A \mathcal{N}})^{i+1}  \frac{1}{N_o \Omega^{\frac{2}{3}}}-2\mathcal{N},
\label{equ:B5}
\end{equation}

This leads to the result $\eta = i/(i+2)$ like that for the complete condensation regime.

\begin{figure}
\caption{ 
Island density versus deposition rates $R$ in our kinetic Monte Carlo
simulations as described in the text. Solid line with open circles: MBE;
dotted line with open squares: SiH with hydrogen evaporation rate $R_e=1200$;
dot-dash line with solid diamonds: SiH with hydrogen evaporation rate $R_e=2000$;
double dot-dash and double dash-dot lines: SiH with H atom deposited randomly onto
available nearest neigbor dangling-bond site and up to next-nearest neighbor sites
respectively;dashed line with solid triangles: two component SiH, SiH$_2$ results.
}
\label{fig1}
\end{figure}

\begin{figure}
\caption{ 
Scaled island distributions for Si MBE for (a) R=1 and (b) R=16. Curves with circles,
diamonds and squares are due to the scaling function (Eq. 3) for i=3,2 and 1 respectively.
}
\label{fig2}
\end{figure}

\begin{figure}
\caption{
Scaled island distributions for two component growth (SiH and SiH$_2$) for (a) R=1
and (b) R=16. Lines with circles, diamonds and squares are as in figure 2.
}
\label{fig3}
\end{figure}

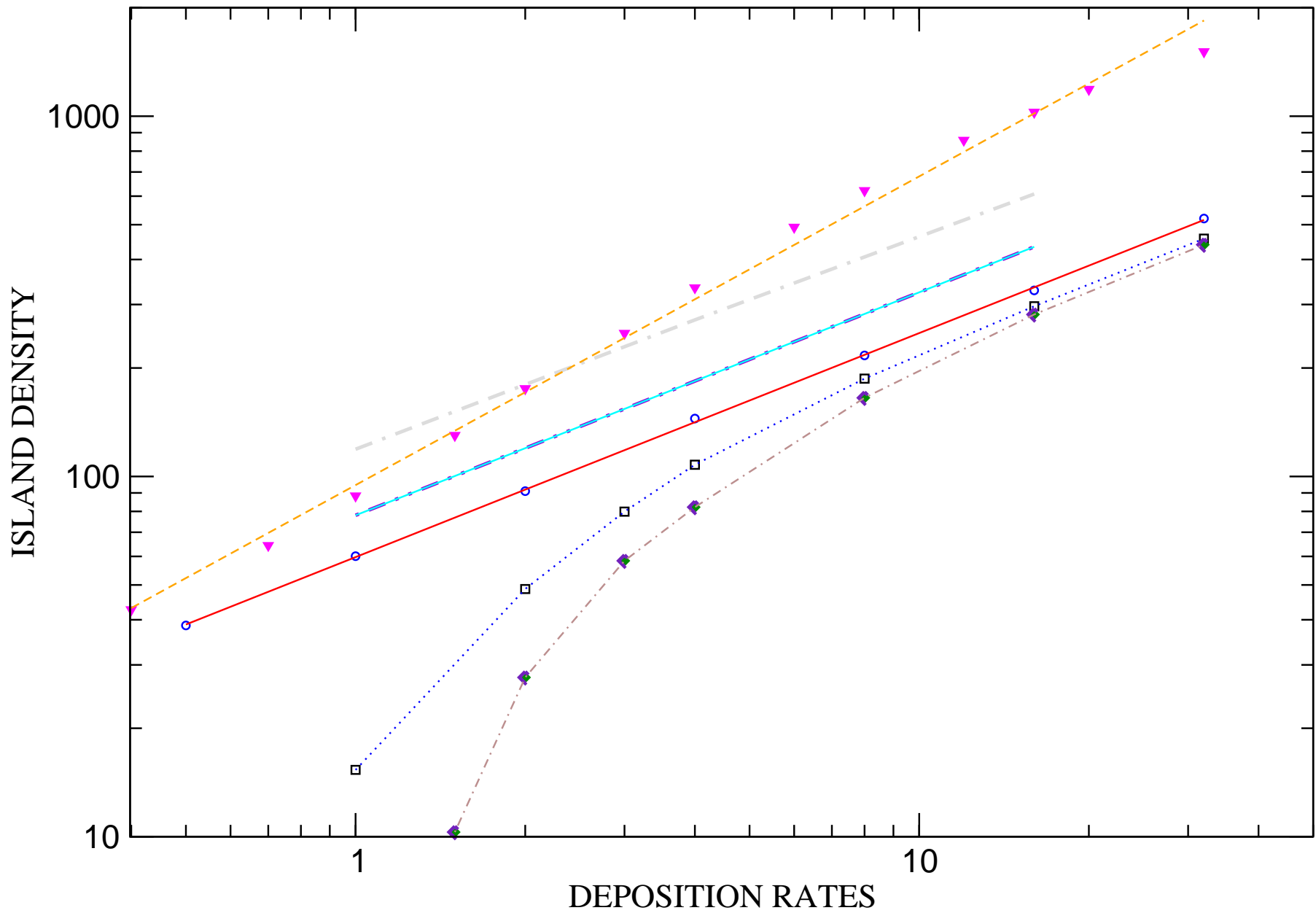
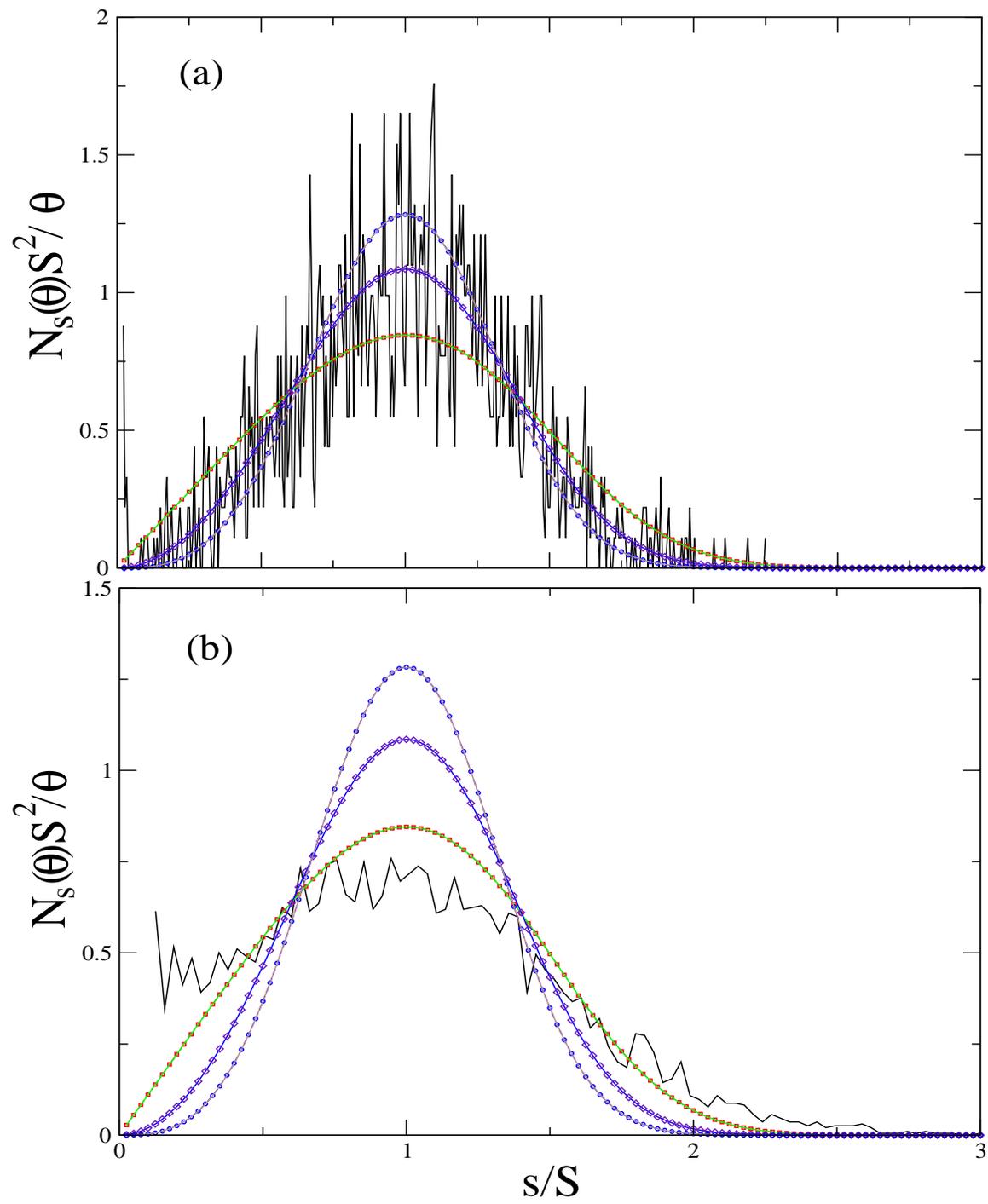
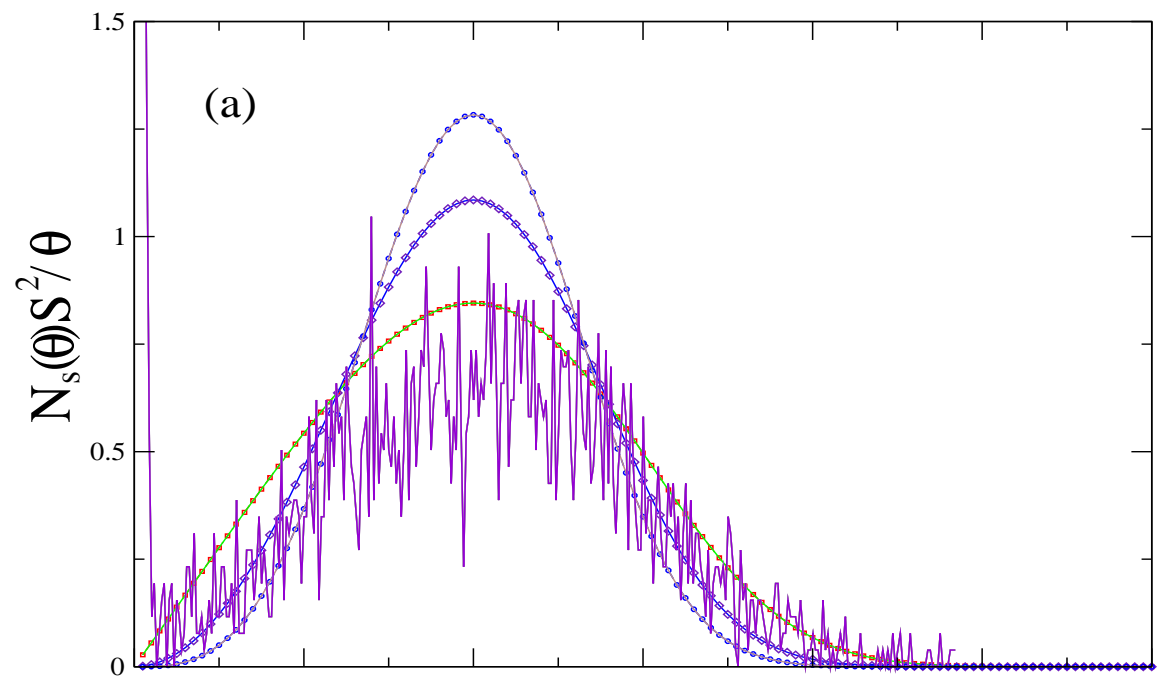
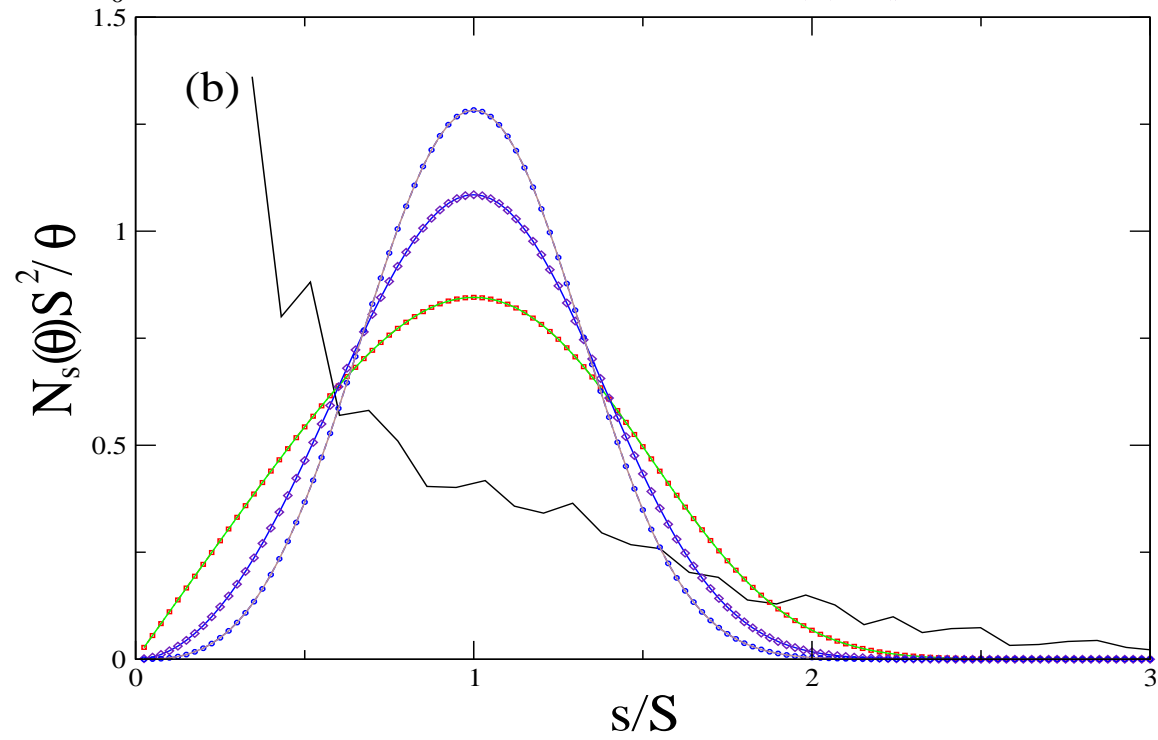
\end{document}